\documentclass[aps,prl,onecolumn,superscriptaddress]{revtex4}

\usepackage{graphicx}
\usepackage{amssymb,amsfonts,amsmath}
\usepackage{appendix}
\usepackage{booktabs}
\usepackage[pdftex,usenames]{color}

\makeatother

\newcommand{\lb}{\label}
\newcommand{\be}{\begin{equation}}
\newcommand{\ee}{\end{equation}}
\newcommand{\av}[1]{\langle #1 \rangle}
\newcommand{\bea}{\begin{eqnarray}}
\newcommand{\eea}{\end{eqnarray}}
\newcommand{\dF}{\mbox{dF}}
\newcommand{\myeqref}{eq$.$~\eqref}
\usepackage{hyperref}
\usepackage{alphalph}
\makeatletter
\newalphalph{\fnsymbolmult}[mult]{\@fnsymbol}{5}
\makeatother

\newcommand{\mycite}[1]{~\cite{#1}}

\begin{document}

\title{Quantifying the effect of temporal resolution on time-varying networks}

\author{Bruno Ribeiro$^{1}$, Nicola Perra$^{2}$, Andrea Baronchelli$^3$\\
~\\
\small{${}^{1}$School of Computer Science, Carnegie Mellon University,\\
 5000 Forbes Avenue, Pittsburgh PA 15213, USA\\
${}^{2}$Laboratory for the Modeling of Biological and
  Socio-technical Systems, \\ Northeastern University, Boston MA 02115,
  USA\\
  ${}^{3}$Department of Mathematics, City University London, \\Northampton Square, London EC1V 0HB, UK}
}

\begin{abstract}
Time-varying networks describe a wide array of systems whose constituents and interactions evolve over time. They are defined by an ordered stream of interactions between nodes, yet they are often represented in terms of a sequence of static networks, each aggregating all edges and nodes present in a time interval of size $\Delta t$. In this work we quantify the impact of an arbitrary $\Delta t$ on the description of a dynamical process taking place upon a time-varying network. We focus on the elementary random walk, and put forth a simple mathematical framework that well describes the behavior observed on real datasets. The analytical description of the bias introduced by time integrating techniques represents a step forward in the correct characterization of dynamical processes on time-varying graphs.
\end{abstract}

\maketitle

Time-varying networks are ubiquitous. Examples are found in the social, cognitive, technological and ecological domains as well as in many others\mycite{holme11-1}. The temporal nature of such systems has a deep influence on dynamical processes occurring on top of them\mycite{morris93-1,morris07-1,clauset07,alex12-1,Rocha:2010,Isella:2011,Stehle:2011nx,Karsai:2011,Miritello:2011,dynnetkaski2011, albert2011sync,Parshani:2010,Bajardi:2011, consensus_temporal_nrets_2012,starnini_rw_temp_nets,perra12-1,perra12-2,lambiotte12-1,Krings:2012cl,Holme:2013ws}. Indeed, the spreading of sexual transmitted diseases, the diffusion of topics over social networks, and the propagation of ideas in scientific environments are affected by duration, sequence, and concurrency of contacts\mycite{morris93-1,clauset07,butts08-1,toro_2007,perra12-1,perra12-2,lambiotte12-1}. In all these cases the timescale characterizing the evolution of the network is comparable with the timescale ruling the unfolding of the process, and they cannot be decoupled.  However, empirical datasets are often reduced to a series of static networks by introducing a time-integrating window, $\Delta t$\mycite{maity2012opinion,carley03-1,rosvall10-1,holme03-1,holme11-1}. This is the case, for instance, of face-to-face interaction networks\mycite{cattuto2010dynamics}, for which the fine-grained temporal resolution of (e.g.) phone call networks is not available, or of infants' semantic networks \mycite{baronchelli2013networks}, whose evolution can be studied only through the analysis of few snapshots\mycite{beckage2011small}. 
In other instances, a time window is introduced to reduce the amount of stored information, or to simplify the application of mathematical frameworks developed for static or annealed systems. This is the case, for example, of online social networks where, although usually the original information has time resolutions down to the second, the available datasets are integrated over different windows of hours, days, months, or even years.
Thus, the introduction of an integrating window is either intrinsic to the system under study or dictated by practical reasons.

In this work we address the impact of an arbitrary $\Delta t$ on the description of a discrete dynamical process taking place upon a time-varying network.
Despite recent results showing that the presence of any level of temporal aggregation may affect the correct characterization of dynamical processes evolving on top of such datasets\mycite{morris93-1,morris07-1,clauset07,alex12-1,Rocha:2010,Isella:2011,Stehle:2011nx,Karsai:2011,Miritello:2011,dynnetkaski2011, albert2011sync,Parshani:2010,Bajardi:2011, consensus_temporal_nrets_2012,starnini_rw_temp_nets,perra12-1,perra12-2,lambiotte12-1,Krings:2012cl,Holme:2013ws},
an analytical  formalization, characterization, and understanding of these effects for a general $\Delta t$ is still missing. 

In particular, we focus on the prototypical random walk process evolving on time-varying networks integrated over a general time window $\Delta t$.
First, we clarify the relevance of the integrating window issue by studying the behavior of random walk processes on real time-varying networks as a function of $\Delta t$. Then, we introduce a mathematical framework that well describes the observed behavior on synthetic activity driven networks\mycite{perra12-1} as well as on two different real datasets.

\section*{Results}

We aim to understand how $\Delta t$ affects the behavior of dynamical processes taking place on time-varying networks. To this end, we consider the fundamental random walk (RW) process 
on two different real time-varying networks in which the links have been integrated over different integrating windows $\Delta t$ (see Fig.~\ref{figure1}). Typically, the RW asymptotic occupation probability $\rho$ (see Methods for the formal definition) is computed grouping the nodes according to their the degree $k$\mycite{noh04,newman10-1,barrat08-1}. The quantity $\rho_k$ is then defined as the average asymptotic occupation probability of a node in the degree class $k$\mycite{noh04,newman10-1,barrat08-1}. However, in time-varying networks the degree of a node is not univocally defined and, more importantly, is a function of $\Delta t$. For example, the degree might be the number of connections integrated over the time window, or the average number of connections across the $T/\Delta t$ static frames (where $T$ is the total time span of the data).  Thus,   the same node could contribute to different degree classes depending on the value of $\Delta t$. We, therefore, focus on a different node measure that has been shown to be mostly invariant to $\Delta t$, namely the activity rate $a$ of a node\mycite{perra12-1}. The activity rate $a$ is  defined as the average rate at which each node interacts with others during the observation period $[0,T]$, and can be interpreted as the intrinsic attitude of each node to engage in interactions with other nodes. We aim to calculate the occupation probability as a function of $a$.

\begin{figure*}
 \includegraphics[width=0.6\textwidth] {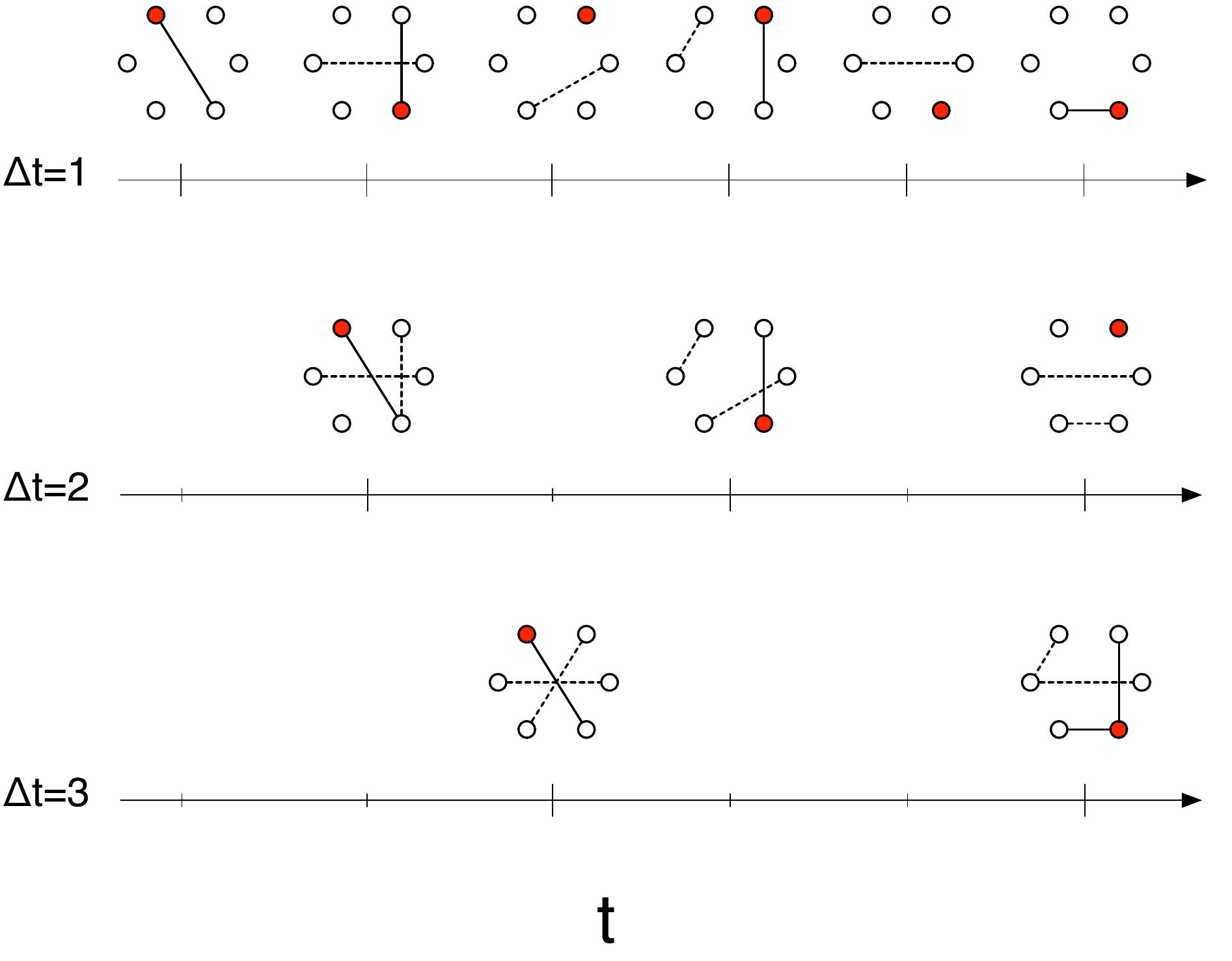}
\caption{Example of time integration on time-varying networks. The random walker is located on the colored node, and can travel on the links depicted as continuous line, while $\Delta t$ defines the integration window. Dashed lines represent links that are present in the system, but are out of reach for the walker.}
\lb{figure1}
\end{figure*}

In our simulations we consider two real time-varying networks, and investigate the RW occupation probability function of activity rate $a$ and the integrating window $\Delta t$:  $\rho_a(\Delta t)$. The first dataset is the co-authorship network of the Physical Review Letters (PRL) journal from 1980 to 2006\mycite{aps10-1}. 
The second dataset is the Yahoo!\ music dataset with  $\sim 4.6 \times 10^5$ songs rated by $ \sim 2\times10^4$ Yahoo!\ users over six months\mycite{Yahoo}. 
We run the RW process over these two time-varying networks for different values of $\Delta t$, and record the occupation probability over multiple runs (see SI for details).
Fig.~\ref{PRL} shows the empirical values of $\rho_a(\Delta t)$ (solid points) observed in the PRL dataset for four distinct values of $\Delta t = \{ 1,10,60,182\}$ days. Error bars represent the the standard deviation obtained from distinct simulation runs starting at times $t_0 \in \{0,1,\dots,\Delta t -1\}$ from the beginning of the dataset. 

The effect of $\Delta t$ is dramatic. 
Over large values of $\Delta t$ the RW behaves roughly as could be expected. The share of random walkers increases with the node activity, i.e., highly active nodes are collect more walkers at the end of the simulation than nodes with low activity.
However, as $\Delta t$ decreases, more active nodes lose their power to attract walkers and the occupation probability becomes more uniform. 
A similar scenario is observed over the Yahoo!\ dataset over four values of $\Delta t$, namely one second, one hour, six hours, and one day (points in Fig.~\ref{Yahoo}).
In the next section we will see that the reason for this behavior rests solely in the probability that the RW sees no edges when it decides to move, which turns out to be a function of three factors: $\Delta t$, the activity of node the walker resides, and the average node activity in the system. 

\begin{figure*}[!!ht]
\includegraphics[width=0.6\textwidth] {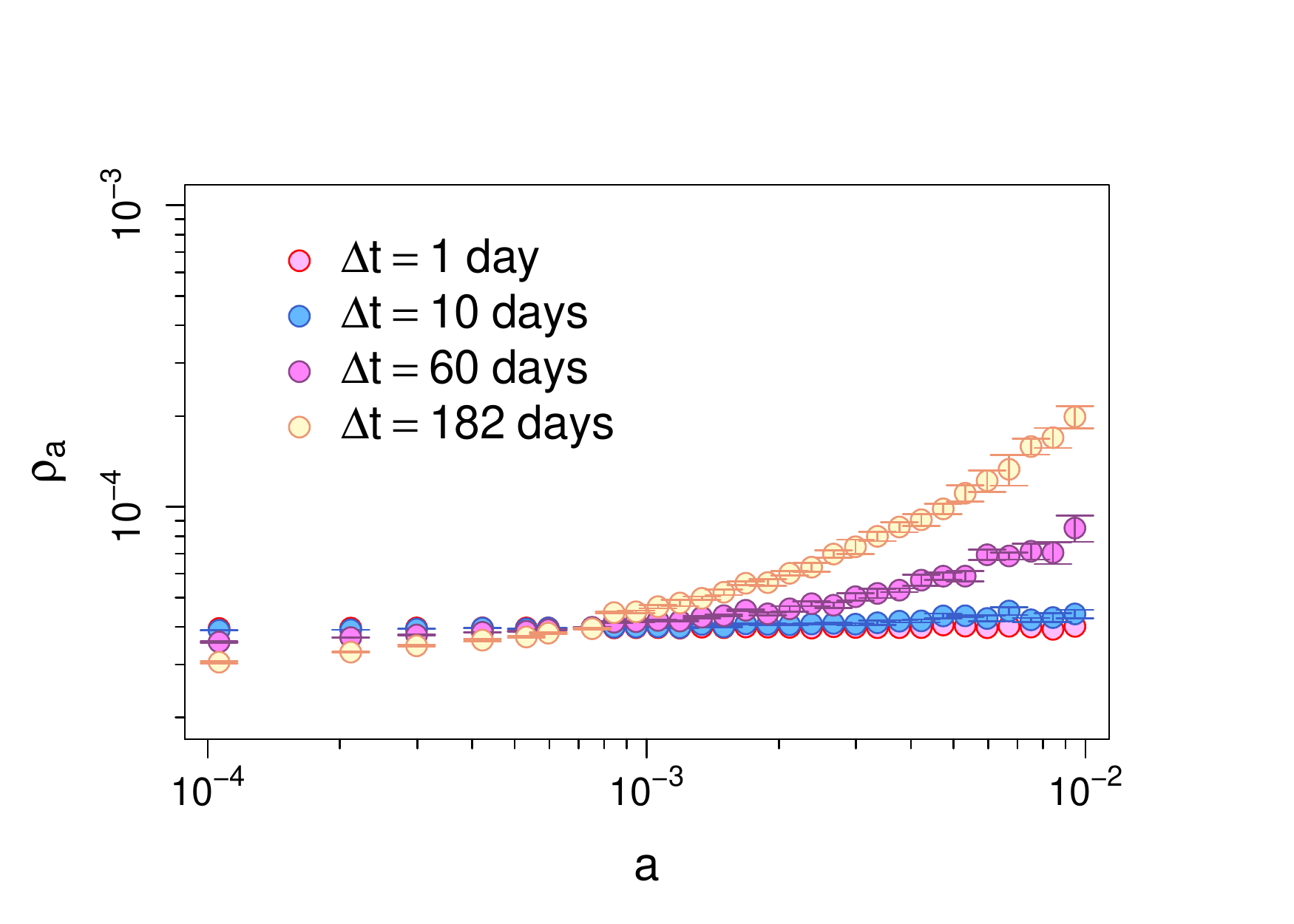}
\caption{Occupation probability $\rho_a$ of a RW at the end of the simulation as a function of node activity. The points are the values of $\rho_a$ of a RW over the Physics Review Letters time-varying co-authorship network from 1980 to 2006 for different integrating windows $\Delta t \in \{1,10,60, 182\}$ days. The error bars are evaluated starting the process at different days from the beginning of the dataset.}
\lb{PRL}
\end{figure*}

\begin{figure*}[!!ht]
\includegraphics[width=0.6\textwidth] {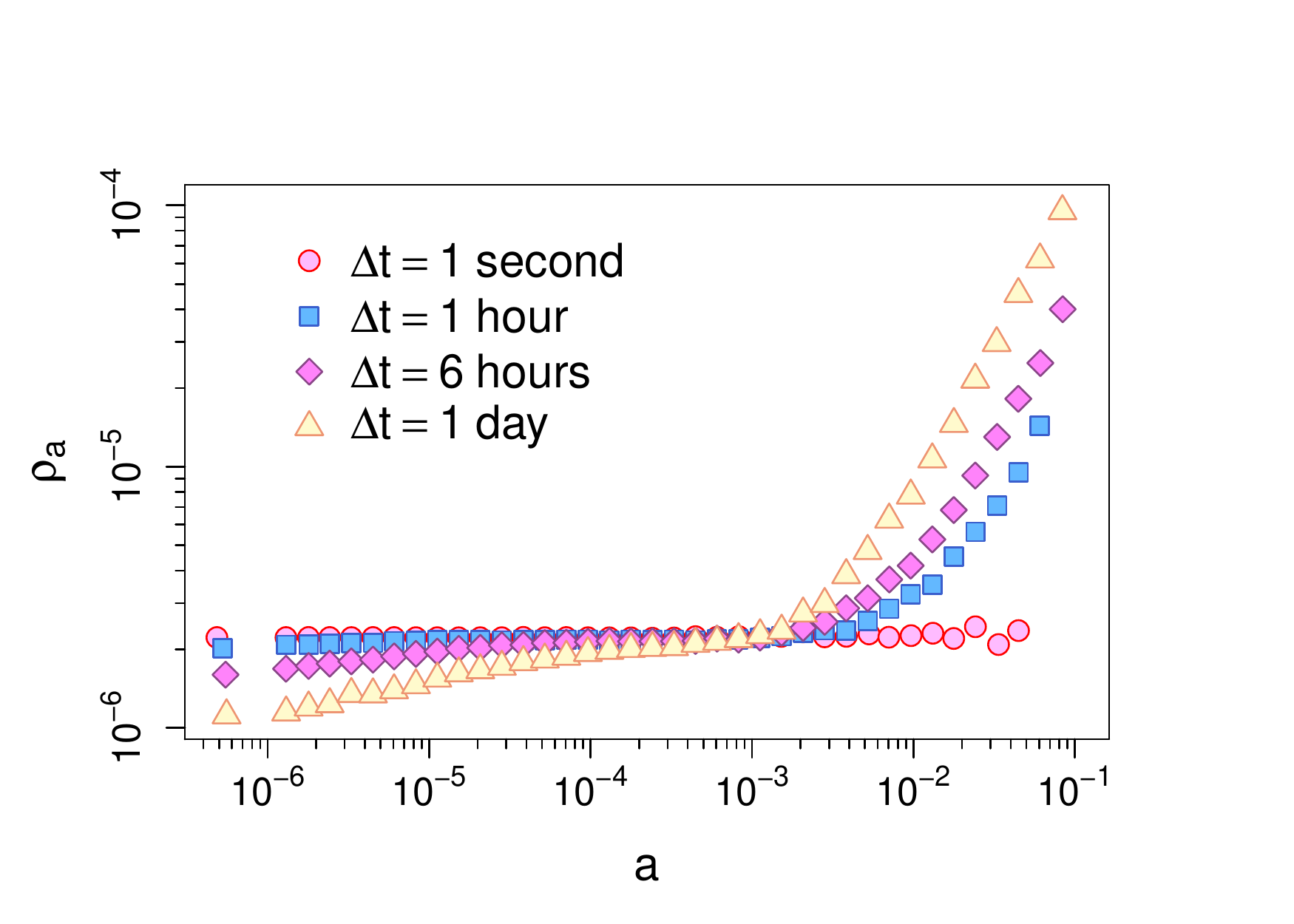}
\caption{Occupation probability $\rho_a$ of a RW at the end of the simulation as a function of node activity. Points represent the $\rho_a$ values of a RW over the time-varying graph of Yahoo!\ song ratings for different integrating windows $\Delta t$ of one second, one hour, six hours, and one day.The standard deviations are too small to be shown in the plots.}
\label{Yahoo}
\end{figure*}


\subsection{Mathematical formulation.}

Let us consider a random walker diffusing at discrete time steps $\Delta t$ over a time-varying network characterized by $N$ nodes. Starting at node $V(t)$ at step $t$, the walker takes step $t+1$ at time $(t+1)\Delta t$ diffusing over a network $G_t(\Delta t)$, where $G_t(\Delta t)$ is the result of the  union of all the edges generated in the interval $[t \Delta t,(t+1)\Delta t)$. We focus on the general case of an arbitrary time aggregation window $\Delta t > 0$. 

We consider a simple class of time-varying networks called activity driven networks~\cite{perra12-1}. The crucial ingredients of these models are: $\mbox{dF(a)}$,
the fraction of nodes with activity rate $a$, and $m$, the number of edges that are simultaneously created by a node (see Methods for further details).
The activity rate determines the probability per unit time for a node to establish ($m$, simultaneously) edges to other nodes in the system. The value of parameter $m$ is dictated by the specific system under consideration.
The case $m > 1$ is appropriate to describe one-to-many interactions, found for example in such systems as Twitter and blog networks\mycite{java07-1,kumar03-1}. On the other hand, $m=1$ describes two-party (dyadic) communications that are characteristic of phone-call and text-message networks\cite{onnela06-1,wu10-1}. At each step $t =0,1,\ldots$ an unweighted network $G_t(\Delta t)$ is generated as follows: 
\begin{enumerate}
\item[a)] $G_t(\Delta t)$ starts with $N$ disconnected nodes; 
\item[b)] The the number of times a node with activity $a$ is {\em active} during interval $\Delta t$, $K_{\Delta t,a}$, is Poisson distributed
\[
\mbox{P}[K_{\Delta t,a}=k]=\frac{\left(a\Delta t\right)^{k}}{k!}\exp(-a\Delta t).
\]
Node generates $m K_{\Delta t,a}$ undirected edges connected to $m K_{\Delta t,a}$ randomly selected nodes (without replacement or self-loops). Inactive nodes in this observed period of $\Delta t$ may receive connections from other active vertices; 
\item[c)] At time $(t+1) \Delta t$ the process starts over from step a) to generate network $G_{t+1}(\Delta t)$. 
\end{enumerate}

Although activity driven networks are Markovian (memoryless) and lack of some properties observed in real temporal systems, they can be
considered as the simplest yet nontrivial framework to study the concurrence of changes in connectivity pattern of the network and dynamical processes unfolding on their structure~\cite{perra12-1,perra12-2}.

To describe the RW behavior, we  need to evaluate the transition probability that a walker starting at a node with activity $a^\prime$  moves to a node with activity $a$ at the next $\Delta t$ time step, $Q_{a|a^\prime}(\Delta t)$. 
Without loss of generality in what follows we focus on the case $m=1$.
Detailed results for the $m>1$ one-to-many interactions are discussed in the Supplementary Information. 
At step $t+1$ the neighbors of $V(t)$ can be classified into two types: 
\begin{enumerate}
\item \textit{Passive destinations}, are neighbors of $V(t)$ connected by edges created due to the activity of $V(t)$ itself. They are randomly selected from the graph and thus their activity is distributed according to  $\dF(a)$. We define $K_{\Delta t,A(t)}$ to be the number of such passive destinations, where $A(t)$ is the activity rate of node $V(t)$.
\item \textit{Active destinations}, are neighbors of $V(t)$ connected to $V(t)$ by edges created due to their own activity. Thus, their activity is distributed as $a \dF(a)/\av{a}$, where $\av{a}$ is the average activity rate in the system. We define define $H_{\Delta t}$ as the number of such active destinations. 
\end{enumerate}
The word \textit{destinations} highlights the fact that the walker moves from $V(t)$ to one of these $K_{\Delta t,a^\prime} + H_{\Delta t}$ neighbors of $V(t)$.
For sufficiently large $N$, $H_{\Delta t}$ and $K_{\Delta t,a^\prime}$ are both Poisson distributed with average $\av{a} \Delta t$ and 
$a^\prime \Delta t$, respectively.
If $V(t)$ has at least one edge, the walker follows the edge of a passive destination with probability $K_{\Delta t,a^\prime} / (K_{\Delta t,a^\prime} + H_{\Delta t})$, while it moves towards an active destination with probability $H_{\Delta t} / (K_{\Delta t,a^\prime} + H_{\Delta t})$.
Unconditioning the latter expressions with respect to the values of $K_{\Delta t,a^\prime}$ and  $H_{\Delta t}$ we obtain 

\begin{equation}
\begin{aligned}
\label{total}
 Q_{a|a^\prime}(\Delta t) =  \Bigg( & \sum_{k=1}^{\infty}\sum_{h=0}^{\infty} \Bigg(\frac{k}{k+h}\dF(a)
 \quad +\frac{h}{k+h}\frac{a\,\dF(a)}{\left\langle a\right\rangle }\Bigg) \frac{(a^\prime \Delta t)^k (\av{a} \Delta t)^h}{k! h!} \\
 & + \sum_{h=1}^{\infty} \frac{a\,\dF(a)}{\left\langle a\right\rangle } \frac{(\av{a} \Delta t)^h}{h!} + \delta(a-a^{\prime})  \Bigg)    \exp(- (a^\prime + \av{a}) \Delta t)  \, ,
\end{aligned}
\end{equation}

\noindent where $\delta(x)$ is the Dirac delta function. While we refer the reader to the SI for the detailed derivation, each term in \myeqref{total} has a simple interpretation. The two terms inside the double sum represent, respectively, the probability that the walker moves to a passive destination that has activity $a$ and the probability that the walker moves to an active destination that has activity $a$. The terms multiplying the two terms inside the double summation are related to the probability that $K_{\Delta t,a^\prime} =  k$ and $H_{\Delta t}=h$. The $\delta(a-a^\prime)$ term considers the probability that the node has no edges after $\Delta t$ and thus the walker must remain at $V(t)$.\\
Thankfully, \myeqref{total} can be simplified (see SI) yielding
\begin{equation}
\label{eq:m1}
  Q_{a|a^\prime}(\Delta t)  = \frac{a^\prime + a}{a^\prime + \av{a}} \dF(a)  (1-\zeta_{a^\prime , \Delta t}) + \delta(a^\prime - a) \zeta_{a^\prime , \Delta t} \, ,
\end{equation}
where $\zeta_{a^\prime , \Delta t}= e^{-(a^\prime +  \av{a}) \Delta t}$ is the probability that no edge is created at a node with activity $a^\prime$ during interval $\Delta t$.
Note that in \myeqref{eq:m1} the parameter $\Delta t$ only affects the probability that no edge is created until the next time step.

To find the RW stationary distribution we first note that the RW on the time-varying network is stationary and ergodic (see SI).
Thus, the RW occupation probability $\rho_a$, defined as the probability of finding the walker in a given node of activity $a$, exists and is unique\mycite{ribeiro12-1}.  
The value of $\rho_a$ is the fixed point solution of the following Chapman-Kolmorogov set of equations\mycite{feller-vol-2}
\begin{equation} \label{eq:rhoa}
\rho_{a} = \frac{1}{N\dF(a)}  \int_{a^\prime \in \Omega} Q_{a|a^\prime}(\Delta t) \rho_{a^\prime} \dF(a^\prime) \, , \quad \forall a \in \Omega \, ,
\end{equation}
where $\Omega$ is the set of all activity rates in the system.
The solution to \myeqref{eq:rhoa} can be obtained numerically. 
Interestingly, we can extend \myeqref{eq:rhoa} to consider lazy random walks where the walker moves with probability $p \in (0,1]$ or does not move with probability $1-p$. 
For the lazy walker we just need to replace $Q_{a|a^\prime}(\Delta t)$ in  \myeqref{eq:rhoa} with $Q_{a|a^\prime}(\Delta t)  p + \delta(a^\prime - a) (1-p)$.
A simple algebraic manipulation shows that $\rho_a$ does not change with $p$.
Hence, the steady state of the lazy walker for any $p \in (0,1)$ is the same as the walker that moves with probability $p = 1$.

We also find that closed-form solutions of~\myeqref{eq:rhoa} exist in the limits of $\Delta t \gg 1$ and $\Delta t \ll 1$.
In the  $\Delta t \gg 1$ case, links are integrated over a large time window and the time-varying network can be considered static. 
Recall that $\zeta_{a,\Delta t}  = e^{(a + \av{a})\Delta t}$.
For $\Delta t \gg 1$ the value of $ \zeta_{a,\Delta t} \approx 0$, $\forall a \in \Omega$, and thus  the second term of \myeqref{eq:m1} is close to zero.
In this scenario $Q_{a|a^\prime}(\Delta t) = C \, (a+\av{a}) \dF(a) $, where $C = 1/2\av{a}$ yielding 
the fixed point solution of \myeqref{eq:rhoa}
\begin{equation} \label{eq:rhoam1}
     \rho_a  \approx \frac{a + \av{a}}{2 N \av{a}} \, .
\end{equation}
The asymptotic occupation probability of a given node of class $a$ is simply proportional to its activity.  Since in the regime of large $\Delta t$ the degree of a node $v$, $k_v$, is proportional to its activity, $a_v$, \myeqref{eq:rhoam1} yields $\rho_{a_v} \propto k_v$. Thus, for sufficiently large $\Delta t$, we recover the well-known behavior of static networks, where the occupation probability of a node is proportional to its degree \mycite{noh04} . Furthermore, in the SI we show that eqs. \eqref{eq:m1}, \eqref{eq:rhoa}, and \eqref{eq:rhoam1} hold for weighted aggregation procedures where integrated edges have weights proportional to how often they appeared during an interval $\Delta t$.\\

In the regime of very short aggregating windows we have 
$
\lim_{\Delta t \to 0} \zeta_{a,\Delta t} \to 1$, $ \forall a \in \Omega.$
Thus, the first term of~\myeqref{eq:m1} is zero yielding $Q_{a|a^\prime}(\Delta t) = \dF(a)$ and the trivial fixed point solution of \myeqref{eq:rhoa} 
\begin{equation}\label{eq:rhouni}
     \rho_a  \approx \frac{1}{N} \, .
\end{equation}
Thus, the walker is equally likely to be found at any node regardless of its activity rate.
In fact, when $\Delta t$ is small the probability a node has more than one edge is close to zero. Consequently, highly active nodes lose and gain walkers at the same rate, giving rise to homogeneous occupation probabilities in \myeqref{eq:rhouni}.
Interestingly, in previous work on general time-varying network processes we show that the result in \myeqref{eq:rhouni} holds even when aggregated snapshots have arbitrary strong spatio-temporal correlations\mycite{ribeiro12-1}.

\subsection{Numerical validation on synthetic networks.} 
\begin{figure*}[!!ht]
\includegraphics[width=0.6\textwidth] {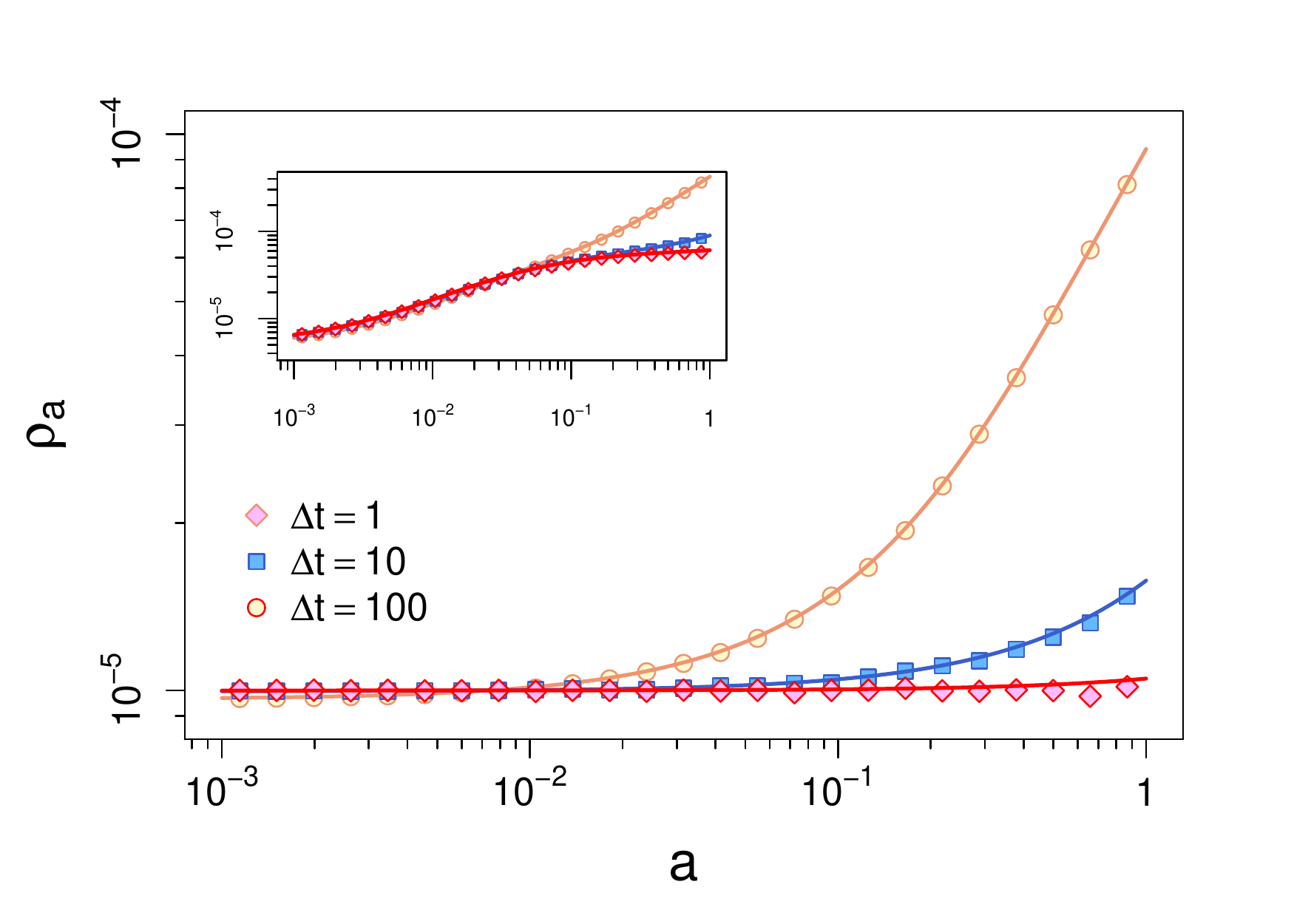}
\caption{Occupation probability $\rho_a$ of a RW over an activity-driven network with activity distribution $dF(a) \propto a^{-2}$, $a \in (10^{-3},1)$, $N = 10^5$, for different values of $m$. Curves in the main plot concern the $m=1$ case, where each node can only simultaneously connect to one node. In the inset, the case $m=6$ is considered, where a node simultaneously connect to six other nodes. Solid curves represent the analytical prediction of~\myeqref{eq:rhoa} integrated over $\Delta t = 1,10,100$ (diamonds, squares and circles) time windows. Note that in both panels as $\Delta t$ gets larger $\rho_a \approx a$. Averages performed over $10^3$ independent simulations.
\label{figure2}}
\end{figure*}

We validated our analytical results through extensive numerical simulations.
We considered networks with $N = 10^5$ nodes and a power-law activity distribution $dF(a) \propto a^{-\gamma}$ (as observed in many real networks\mycite{perra12-1}), restricted to the interval $ \Omega = [10^{-3},1]$ to avoid divergencies in the limit $a \ll 1$. As shown in Fig.~\ref{figure2}, the exact solution reproduces the simulations accurately for the entire spectrum of integrating windows $\Delta t$ (case $m=1$ in main panel).
Interestingly, as $\Delta t$ grows, the occupation probability increases sharply in high-activity vertices while slightly decreasing at low activity nodes.
Moreover, as $\Delta t $ increases $\rho_a  \propto a$ as predicted by \myeqref{eq:rhoam1}, while
as $\Delta t $ gets smaller, $\rho_a = 1/N$, as predicted by \myeqref{eq:rhouni}.
The equations describe correctly also the behavior observed for one-to-many simultaneous connections $m$, characterized by a smoother increase in $\rho_a$ at high activity nodes (see $m=6$ case in Fig.~\ref{figure2}, inset). The SI contains more details on the formulation of the $m>1$ case.

\subsection{Numerical validation on real-world networks.} 

The analytical framework discussed above qualitatively reproduce also the behavior observed in real datasets.
In Figs.~\ref{PRLth} and~\ref{Yahooth} the solid lines show the numerical solution obtained by applying \myeqref{eq:m1} into \myeqref{eq:rhoa}  (see SI), for the PRL and Yahoo!\ datasets, respectively. The gray points in Figs.~\ref{PRLth} and~\ref{Yahooth} reproduce the simulation results already shown in Figs.~\ref{PRL} and~\ref{Yahoo}, respectively.
All numerical solutions use the same activity distribution $\dF(a)$, extracted from the time-varying graph of $\Delta t = 1$ day for the PRL dataset and $\Delta t = 1$ second for the Yahoo!\ dataset ($\dF(a)$ extracted from larger values of $\Delta t$ provide similar results\mycite{perra12-1}, see SI for details).

The theoretical results accurately describe real data, with some deviations for nodes in the intermediate activity range at $\Delta t$ of one day.  The RW occupation probability is uniform and independent of node activity for small $\Delta t $ as predicted by \myeqref{eq:rhouni}.
As predicted by \myeqref{eq:rhoam1}, the RW occupation probability $\rho_a$ approaches $(a + \av{a})/(2 N \av{a})$ (black curve) as $\Delta t$ increases, an effect particularly noticeable for high-activity nodes.
It is also worth highlighting that the data matches well the theoretical equations for the case  $m=1$,
suggesting a connection between the datasets and the fundamental mechanisms described in our model (for the similarity in behavior between $m=1$ and projected networks such as the PRL co-authorship networks see SI).

\begin{figure*}[!!ht]
\includegraphics[width=0.6\textwidth] {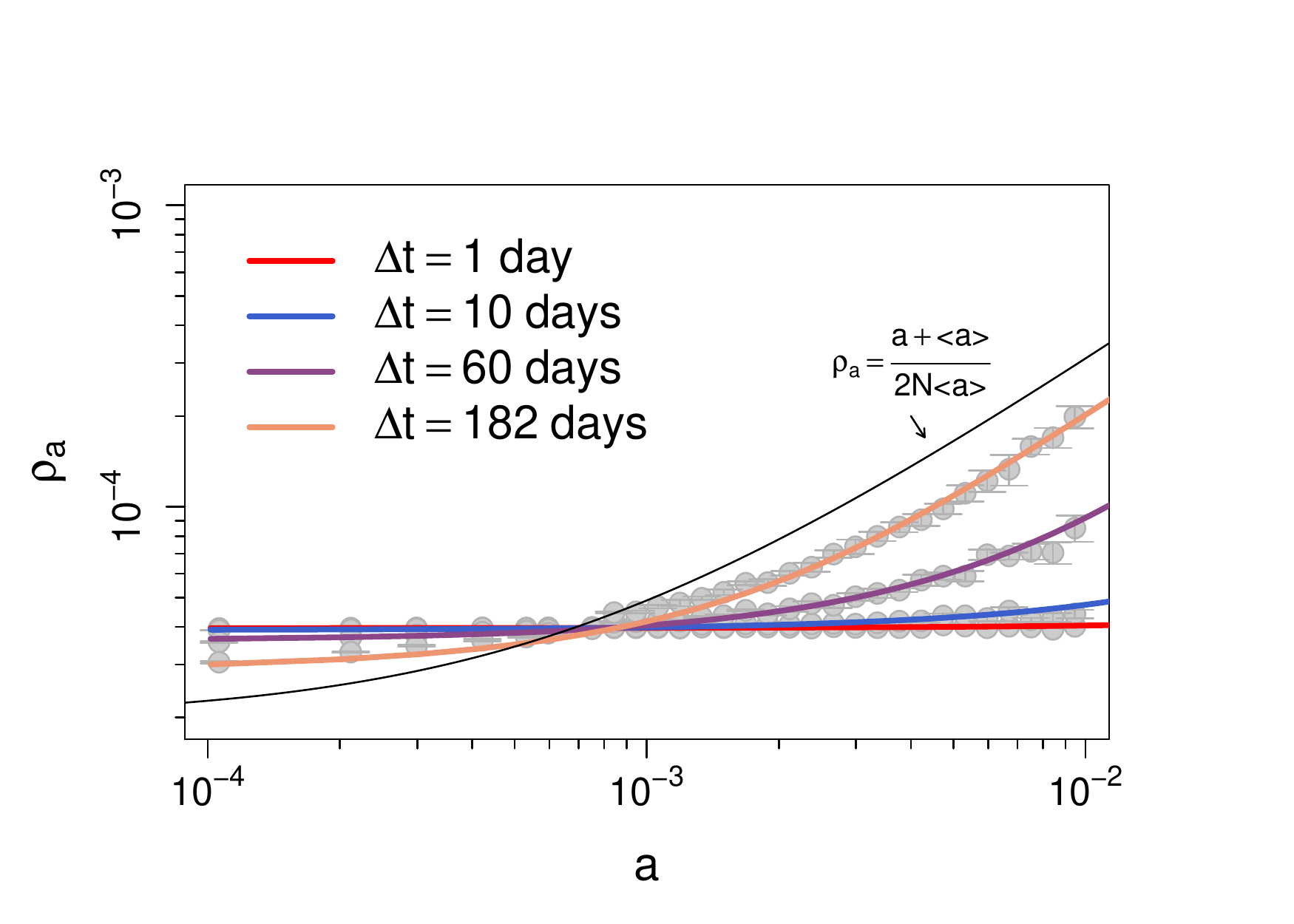}
\caption{Occupation probability $\rho_a$ of a RW at the end of the simulation as a function of node activity. The points are the values of $\rho_a$ of a RW over the Physics Review Letters time-varying co-authorship network from 1980 to 2006 for different integrating windows $\Delta t \in \{1,10,60, 182\}$ days. The solid curves show the respective numerical solutions of \myeqref{eq:rhoa} and the black curve shows \myeqref{eq:rhoam1}.}
\lb{PRLth}
\end{figure*}

\begin{figure*}[!!ht]
\includegraphics[width=0.6\textwidth] {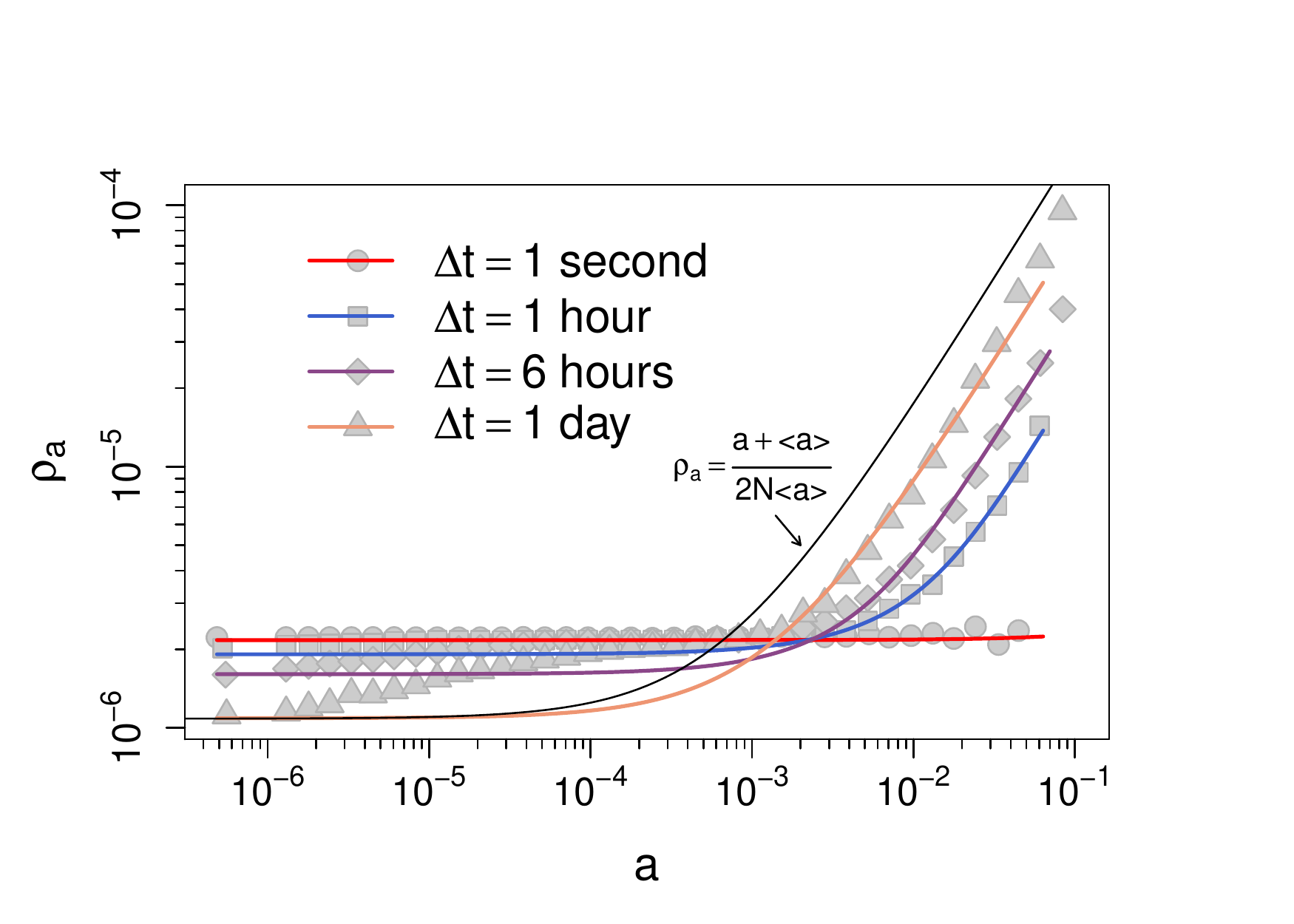}
\caption{Occupation probability $\rho_a$ of a RW at the end of the simulation as a function of node activity. The points are the values of $\rho_a$ of a RW over the time-varying graph of Yahoo!\ song ratings for different integrating windows $\Delta t$ of one second, one hour, six hours, and one day. The solid curves show the respective numerical solutions of \myeqref{eq:rhoa} and the black curve shows \myeqref{eq:rhoam1}. }
\label{Yahooth}
\end{figure*}


\section*{Discussion}
Our results clarify the effect of time aggregation procedures on the behavior of the RW, taken as the simplest instance of dynamical process, even when aggregation windows are ``short''. We have quantified this effect in a rigorous mathematical framework that (i) allows us to recover the results concerning static networks in the limit of infinite aggregation windows, (ii) accurately describes the behavior observed in numerical simulations upon synthetic time-varying networks, and (iii) captures the phenomenology observed on real datasets. Overall, while for practical or technical reasons researchers are often forced, or simply tempted, to work with time aggregated representations of time-varying networks, our work suggests that caution should be used when drawing general conclusions about dynamical processes based upon time-aggregated networks. At the same time, moreover, our theoretical results may help to investigate possible distortions introduced by the aggregating windows of data collection methods. \\
The proposed framework considers inherently discrete processes, such as spreading phenomena in contact networks that are, also at the smallest time resolution possible, discrete. We leave the generalization to continuous processes for further work.

\section*{Methods}

\subsection{Occupation Probability.} 
The asymptotic occupation probability is the steady state probability of finding the walker in a node with activity $a$, which is guaranteed to exist and be unique if the time-varying network that is stationary, ergodic, and T-connected (see SI), such as in activity driven networks. A time-varying network is T-connected if there is a temporal path between any two nodes\mycite{ribeiro12-1}. 
In our simulations we consider the RW occupation probability $\rho_a$ to be the probability of finding the walker in a node with activity $a$ at the end of the simulation period $[0,T]$, given the walker starts at a random node.

\subsection{Activity-driven networks.} Activity-driven network models are based on the activity patterns of nodes, that are used to explicitly model the evolution of the network structure over time\cite{perra12-1}. 
%
%
It can be shown that the full dynamics of the network are encoded in the activity rate distribution, $\dF(a)$ and that the time-aggregated measurement of network connectivity yields a degree distribution that follows the same functional form as the distribution $\dF(a)$ in the limit of small $k/\Delta t$ and $k/N$\mycite{perra12-1}. This is an important feature of the model, that is able to reproduce basic statistical properties found in many real networks giving a simple prescription to characterize explicitly dynamical connectivity patterns. \\

\subsection{Datasets \& Simulation.}

In this study we considered two different empirical projections of bipartite time-varying networks. The collaborations in the journal ÓPhysical Review LettersÓ (PRL) published by the American Physical Society\cite{aps10-1}, and the Yahoo!\ music dataset made available by Yahoo!\cite{Yahoo}.
~\\

\noindent
\emph{PRL dataset}. The bipartite network representation of this dataset has two type of nodes: authors and papers. An author is connected to all the papers she/he wrote in a integrating window $\Delta t$. We study the bipartite projection of the authors. In this representation each author of an article in PRL as a node. Undirected edges connect authors that collaborate in the same article. We focus just on small collaborations filtering out all the articles with more than $10$
authors. We
consider the period between $1958$ and $2006$. The datasets contains $80,\!554$ authors and $66,\!892$ articles. The smallest timescale available is one day. 
\\
\\
\emph{Yahoo!\ music dataset.} In this dataset the bipartite network has two type of nodes: users and songs. We study the bipartite projection over the songs. Each node is a song and two songs are connected if at least one user rated both in a time window $\Delta t$. The dataset  contains $4.6 \times 10^5$ songs rated by $199,\!719$ users of Yahoo!\ users collected in the course of six months\cite{Yahoo}. User activity is recorded at a time resolution of seconds. 

\noindent
\emph{Simulation setup.} We obtain the empirical walker occupation probability, $\rho_a$, as follows.
Construct the transition probability matrix $P_t$ associated to the RW on the $t$-th aggregated network $G_t(\Delta t)$, $t=0,\ldots, \lfloor T/\Delta t \rfloor,$ where $T$ is the time of the last event in the dataset. The empirical RW occupation probability is obtained by multiplying the matrices $P_0 \, P_1 \cdots P_n$ and then left-multiplying the result by the vector $(1/N,\ldots,1/N)$, which gives equal probability that for the walker to start at any node. We note in passing that similar results are obtained when the walker starts at a handful of high activity nodes.

\noindent
{\bf Acknowledgments.}\\
This work was performed while B.R.\ was a postdoctoral researcher at the University of Massachusetts Amherst visiting the MoBs Lab at Northeastern University.
We thank Yahoo!\ and APS for providing the data used in this work. B.R.\ was partially supported by the NSF grant CNS-1065133. B.R.\ and N.P.\ were partially supported by the ARL Cooperative Agreement W911NF-09-2-0053. 
The views and conclusions contained in this document are those of the authors and should not be interpreted as representing official policies, either expressed or implied of the NSF, ARL, or the U.S. Government. The U.S. Government is authorized to reproduce and distribute reprints for Government purposes notwithstanding any copyright notation hereon.
  \\
 {\bf Author Contributions } \\
 B.R, N.P \& A.B designed research, B.R \& N.P performed simulations,  B.R, N.P \& A.B analyzed the data,  B.R derived the analytical results. All authors wrote,
reviewed and approved the manuscript.\\
\\
{\bf Competing financial interests}\\
The authors declare no competing financial interests.

\section*{References}

\end{document}